# Modeling of Key Quality Indicators for End-to-End Network Management: Preparing for 5G

Ana Herrera-García, Sergio Fortes*, Eduardo Baena, Jessica Mendoza, Carlos Baena, Raquel Barco

Dpto. Ingeniería de Comunicaciones, E.T.S.I. Telecomunicación, Universidad de Málaga, Andalucía Tech

29071 Málaga, Spain

{ahg, sfr, ebm, jmr, jcbg, rbm}@ic.uma.es

**Abstract**— Thanks to evolving cellular telecommunication networks, providers can deploy a wide range of services. Soon, 5G mobile networks will be available to handle all types of services and applications for vast numbers of users through their mobile equipment. To effectively manage new 5G systems, end-to-end (E2E) performance analysis and optimization will be key features. However, estimating the end-user experience is not an easy task for network operators. The amount of end-user performance information operators can measure from the network is limited, complicating this approach. Here we explore the calculation of service metrics [known as key quality indicators (KQIs)] from classic low-layer measurements and parameters. We propose a complete machine-learning (ML) modeling framework. This system's low-layer metrics can be applied to measure service-layer performance. To assess the approach, we implemented and evaluated the proposed system on a real cellular network testbed.

## 1 INTRODUCTION

Worldwide mobile subscriptions are expected to exponentially increase in the following years. But not just the number of subscribers is boosting, user expectations are increasing due to the availability of more sophisticated terminals and the large variety of offered services. This growing demand for data services has driven the development of different mobile standards in the last 10 years. 5G is the next generation mobile network which plans to respond to these changes in society and industries. Among 5G network key features are higher data rates (up to 20 Gbps) or 1 ms over-the-air latency. Furthermore, in its conception, the 5G network aims to be a tool to fully cover the diverse requirements of the different *end-to-end* (E2E) applications, achieving the satisfaction of the users by providing them full *quality of experience* (QoE).

Traditionally, the quality of network services has been analysed via *Key Performance Indicators* (KPIs), e.g. throughput, call drop rate or packet latency. In this sense, service providers have typically based their analysis of the network on low-layer non-E2E parameters and performance measurements as they are directly measurable by classic management mechanisms. However, these metrics are not enough to fully optimize and grasp the E2E service performance since they do not consider the service characteristics and therefore, do not reflect consumer perception. For example, it is very common that the network KPIs show a proper performance, but the quality of specific services is not enough. Thus, it is necessary to understand the relationship between network performance and configuration with the services performance and end-user experience.

To fully tackle the E2E point of view other metrics are required, namely application-layer *Key Quality Indicators* (KQIs) [1]. The intention of using KQIs is to better reflect the customer experience by the definition of objective service indicators that directly impact the QoE experienced by the user for each service, e.g. video streaming stall frequency, page download time, initial file transfer delay, etc. Monitoring the KQIs will therefore be a key to properly analyse problematic situations and improve the performance of the network. In this sense, service-oriented analysis is also one of the key features to consider in the management of these new 5G systems.

Following a review of the literature, there have been some initial approaches to this end, particularly for non-cellular Internet provision (e.g. wired broadband connections). In [2] a generic web services mapping is performed to estimate the KPI-KQI contribution to the QoE via machine learning techniques using principles of web engineering and web quality requirements. Referred to voice services, in [3] and [4], also a mathematical mechanism able to estimate the KPI/KQI contribution to QoE metrics is proposed. In [5] an exponential model is used to establish the KPI/KQI relationship for LTE video streaming and voice service. In terms of video service, most of the researches has focused on identifying the most relevant KQIs and studying their impact on user perceived video quality, the MOS (Mean opinion score). [6] proposes a machine learning method to estimate the bitrate of HTTPS YouTube video streaming and studies the impact of deviations in bitrate on the KPI parameters used in MOS. In [7] a model is proposed to estimate the MOS based on KQI metrics using the Analytic Hierarchic Process algorithm. Reference [8] studies the impact of radio access parameters in the MOS. The work in [9] studies YouTube and Netflix video traffic over mobile networks and shows that the network bandwidth and the CPU computing power impact on video QoE.

However, the vast majority of these studies are focused on the contribution of the network and service performance to the customer experience [2]-[3]-[4]-[5]-[8]-[9], and not in the relationship between network performance and configurations parameters with the service performance. In addition, they are not adapted to the specific case of cellular networks and in concrete [2]-[3]-[6]-[7], to the RAN (Radio Access Network), characterized, among others, by changing conditions.

The present work proposes a novel ML-based service-agnostic methodology for KQI prediction based on cellular low-layer performance metrics and configuration parameters. This system is implemented and evaluated based on experiments conducted in a real LTE network.

In this way the remainder of this paper is structured as follows. Section II deepens in the motivation of this work analysing the monitoring conditions and existing approaches related to cellular networks. The proposed methodology for the prediction from low-layer metrics to KQIs is outlined in Section III. In Section IV, the evaluation scheme, carried out in an LTE network testbed, is described together with its results. Finally, in Section VI, the conclusions are drawn.

## 2 MOTIVATION

5G networks are expected to cover a large variety of devices/applications, scenarios and use case verticals, all with strongly heterogeneous services requirements. To cope with this, the 3GPP has put the focus on the service KQIs as a way to grasp E2E service performance and the user quality of experience and guide the OAM (Operation, Administration and Management) activities. In this way, several services such as file transfer, video streaming, web-browsing are characterized in terms of different KQIs [10]

The resulting E2E paradigm implies an important change in the inputs that the optimisation techniques are based on. Traditionally, the focus of the optimizations has been based on improving low-layer network performance metrics. Conversely, new mechanisms aim to improve specific service metrics [11].

However, this approach has important challenges due to the difficulty of measuring the KQIs in real scenarios [6] [12]. For example, HTTP has been widely used in many applications especially for video transmission or web. While in the past, the services KQIs could be derived via deep packet inspection of these higher-layer protocol messages, users and content providers are becoming increasingly aware of the security and confidentiality risks and hence the introduction of encryption in their services has become widely implemented. In this way, HTTP over secure socket layer (HTTPS) has been adopted in most of the services. In these cases, bitrate or other specific service KQIs can no longer be directly obtained from the traffic analysis. This encryption mechanism makes difficult to compute KQIs since more sophisticated and complex applications are needed by the operator to get the information about the service performance.

This limitation adds extra complexity to the measurements campaigns. Regarding client-side applications, although they could provide accurate views on several indicators, they are not under the operator's control. When operators' monitoring applications might be used in the user equipment (UEs), the access to the performance of other applications in execution is still limited. For example, they do not typically have access to the services logs or do not know about the packet payload formats required to compute the estimated service metrics for these applications [13]. Also, the level of intrusiveness to the end users made the use of application-based monitoring very limited.

UE traces gathered from the network side are also typically restricted to the layers associated to the radio technology (below IP), suffering also the indicated restrictions in case of encrypted traffic. Advance billing mechanisms such as Event Data Records (EDR), although can provide information of the utilization of different applications, do not contain the necessary level of detail to estimate service performance.

In this way, drive tests remain in fact the main tool able to provide the operator with monitoring information at UE-level and with application-layer granularity. However, both traces and drive tests imply huge costs and time demands, requiring large campaigns to ensure the different services to be relevantly characterized under different conditions. This makes also not advantageous to use them continuously during network operation as an input to OAM activities.

## 3 METHODOLOGY FOR KPI TO KQI PREDICTION

The presented challenges to gather service-level KQI from the cellular network highlight the need for mechanisms to allow their continuous estimation from low layer available data. Therefore, a framework to implement such prediction mechanism is proposed and then implemented.

To do so, the service KQIs may be determined by mathematical combinations of configuration parameters and performance metrics (e.g. KPIs) through complex mapping. Here, each of the considered low-layer inputs might impact KQIs differently depending on the service. Thereby, a flexible mechanism able to decide their contribution according to the service is necessary. This can be solved using Artificial Intelligence and, specifically, Machine Learning (ML) techniques, as capable of generalizing behaviours from information provided as training.

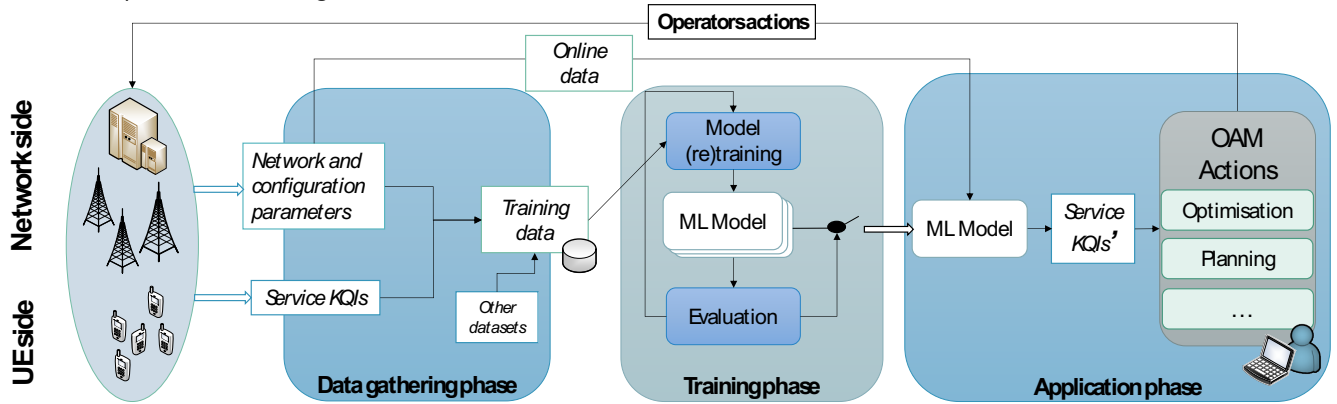

*Figure 1 KQI modelling and prediction framework*

To the purpose of this work, supervised regression is chosen as the baseline approach to implement this block. Regression analysis is a common statistical technique used to investigate the relationship between variables as it is widely used in multiple areas. This group of techniques predicts the outcome of *dependent* variables from *independent* variables (also called *predictors*). In this paper, the service metrics (KQIs), the dependent variables, are predicted based on low-layer directly-measurable network metrics (configuration parameters and KPIs), the independent variables.

Figure 1 describes the proposed framework used to assess the KQI prediction from low-layer metrics. This framework encompasses three main distinct processes: the data gathering phase, the training phase and the application phase.

As input of the system, a measurement set is needed. This data must be composed by service metrics (KQIs) and network metrics (configuration parameters, KPIs, etc.). This data collection can be performed as massive measurement campaigns with no intelligence or using design of experiment (DOE) techniques to cover efficiently all the possible situations. Although the acquisition of this data has the same challenges as described in the previous section, the objective of the framework is to base all future KQI estimation only in a pre-existent KQI dataset, without having to measure the high-level metrics continuously under the operation of the network. Also of great relevance would be the level of completeness and statistic relevance of the training data, and it is going to be further analysed in the evaluation section.

In the training phase, this data serves as input for the construction of the model, linking the KQIs with the low-layer metrics. Before the generation of the model, feature selection techniques might be applied to determine, for each KQI, which are the most impacting low-layer metrics that should be considered for the model construction. Also, feature extraction techniques could be applicable, although it implies a level of abstraction with respect to the input values that it is sometimes not desirable, especially when human operators are involved in the monitoring of the system [14].

Based therefore in the selected data, the model is constructed and evaluated to determine its fitness. This is performed following a classical k-fold cross-validation. The evaluation is done using a set of low-layer metrics and parameters as test data and comparing the estimated KQI (**KQI'**) with its measured values (**KQI**). To evaluate the quality of the models, the coefficient of determination, denoted by $R^2$, is used. This coefficient expresses the proportion of the variance for a dependent variable that's explained by an independent variable and it is a common measurement on how well the model matches the data. An $R^2$ value of 1.0 reveals that the model perfectly fits the data, while a negative or close to zero value expresses an improper modelling or the lack of dependency of the modelled variable in respect to the used inputs.

If the resulting model is adequate, it is ready to be applied for the online KQIs prediction. Otherwise, it might be recalculated with additional data. In the application phase, the model predicts the KQIs based solely on low-layer metrics. In this situation, the model takes as input data configuration and performance of the network and obtain the predicted service performance KQIs for these conditions. During this phase, the framework also incorporates the possibility of adding additionally acquired datasets to the existing ones to re-train the available models.

This proposed approach is relevant for several reasons. Firstly, it allows to use the dataset gathered from time-constrained measurements (such as drive tests, UE traces or end-user app-based campaigns) to train models for future OAM activities. It also allows to predict the KQIs for different services even when no user is making use of them at a specific moment. In this way, this framework also allows the operator to optimise its network by tuning the configuration parameters that have greater influence on the specific service or even anticipate the performance of services that are not currently running in the network and therefore, to adapt the network to the new scenario. This way, an easy and fast adjustment can be done to provide the best quality to the end user.

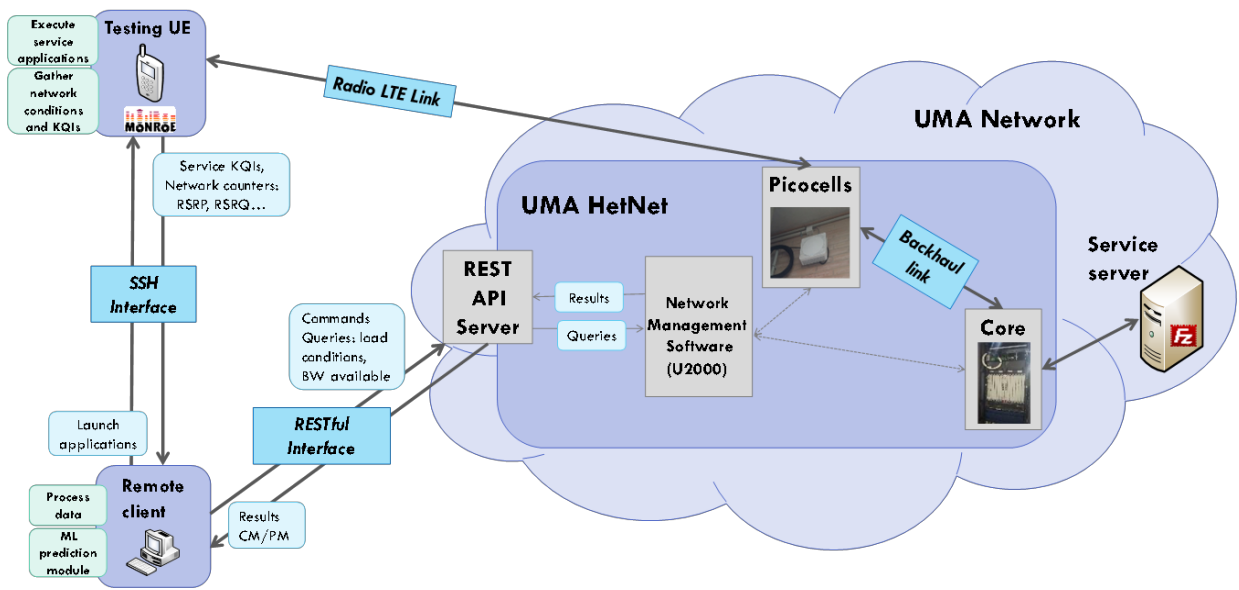

*Figure 2 Prediction framework architecture.*

## 4 EVALUATION

### 4.1 Set-up

In order the collect statistics from the different services, a set of measurements campaigns have been performed in the University of Málaga Heterogenous Network (UMAHetNet) testbed shown in Figure 2.

The UMAHetNet is a full indoor LTE network deployed in the ETSI de Telecomunicación (Telecommunication Engineering School) of the University of Málaga. It is composed by 12 LTE-WiFi Huawei BTS3911B picocells and the core network elements. Both the picocells and the core are fully configurable.

The measurements campaign involves one of these picocells and a static testing UE connected to the network. This testing UE has been developed in the scope of the European H2020 Research Project MONROE - Measuring Mobile Broadband Networks in Europe. It runs a Linux software (in concrete, a SMP Debian 4.9.88-1+deb9u1 x86_64 GNU/Linux 4.9.0-6-amd64) and is based on a PC Engines APU2D4 motherboard equipped with an LTE cat6 modem.

To the purpose of this work, this testing UE is responsible of two main tasks, firstly, of gathering the network radio conditions, and secondly, of executing the different service experiments and gather the experimented KQIs. About the first point, this is performed by a script that collects, using a client-server socket, the radio conditions from the LTE modem: the RSSI (Received Signal Strength Indicator), RSRP (Reference signal received power) and RSRQ (Reference Signal Received Quality). About the second task, each service has its own requirements and thus, its own service KQIs. For this evaluation, the file transfer service has been analysed and the File Transfer Protocol (FTP) has been used as the standard network protocol used for the transfer of files between a client and server on a network. The 3GPP has defined in [10] the KQIs which reflect the end-to-end performance of this service: the initial file transfer delay (s), the file transfer average throughput (Mbps) and the total file transfer delay (s). The **initial file transfer delay** refers to the period from the connection to the server to the moment the very first content is received. The **file transfer average throughput** describes the average data transfer rate measured throughout the entire connection. The **total file transfer delay** is the whole time needed to complete a successful transmission.

The CURL library (https://curl.haxx.se/) has been used to download the files by FTP and collect the service metrics. The metrics provided by this tool that have been considered in the experiments are the average download speed, the total time and setup time. They correspond to the average throughput, the total transfer delay and the initial transfer delay defined for the file transfer service respectively.

Once the experiments have been executed, the ML prediction module processes the results and builds the models. This module is executed in a remote client which is connected to the UE via an SSH interface. Through it, the script receives the experiments results. These results have information about the KQIs under specific radio network conditions and network configuration. Also, this script, is responsible for tuning the network parameters, such the available bandwidth in the cell, to create a complete training dataset under different configuration conditions, which are also stored.

The interaction between the ML prediction module and the network is performed by a REST server. The implemented RESTful interface gives access to configuration functions using MML (Man-Machine Language) commands. This REST API Server sends the commands to the network management software and changes the configuration parameters. Finally, also an FTP server was configured to host the files to download. For that, FileZilla has been chosen, being the server configured to grant the maximum available resources.

**4.2 Prediction implementation**

For the implementation and assessment of the proposed KQI prediction framework, an initial set of low-layer indicators have been considered: the RSRP, RSRQ, RSSI from the UE side and the available bandwidth and the current network load from the network side. These are metrics relatively easy to collect by classical management systems and therefore are used as the independent variables or inputs of the model. Conversely, the KQI data captured by the testing UE are considered the dependent variables of the models to be constructed. To be able to estimate the KQIs for different user demanded files, their size is also included as one of the independent variables.

In this way, 9000 samples have been recorded of different file sizes (1KB, 10KB, 100KB, 500KB, 1MB, 5MB, 10MB, 20MB and 100MB), four different picocell bandwidths (5, 10, 15 and 20 MHz) and three distinct demand conditions. The latest represent different resource cell traffic demand scenarios under which the user could download the files. The three levels are: no additional load in the network, low and medium load. Both low-layer metrics and KQIs have been gathered for all possible combinations of these parameters.

As applicable solution to implement the ML models, five main regression techniques have been evaluated for the training phase: *linear regression* (LR), *step wise linear regression* (SW-LR), *support vector regression* (SVR), *decision tree regression* (DTR) and *random forest regression* (RFR) [15]. LR is the most basic regression technique and models the linear relationship between a dependent variable and the independent variables. SW-LR evolves from LR by implementing a sequential removal of the weakest correlated independent variables from the model. SVR is an adaptation of Support Vector Machines (SVMs) that it is trained using a symmetrical loss function which equally penalizes high and low misestimates. DTR predicts responses to data following the decisions in a tree from the beginning node to the leaf node which contains the response. Finally, RFR is a predictive model composed of a weighted combination of multiple regression trees.

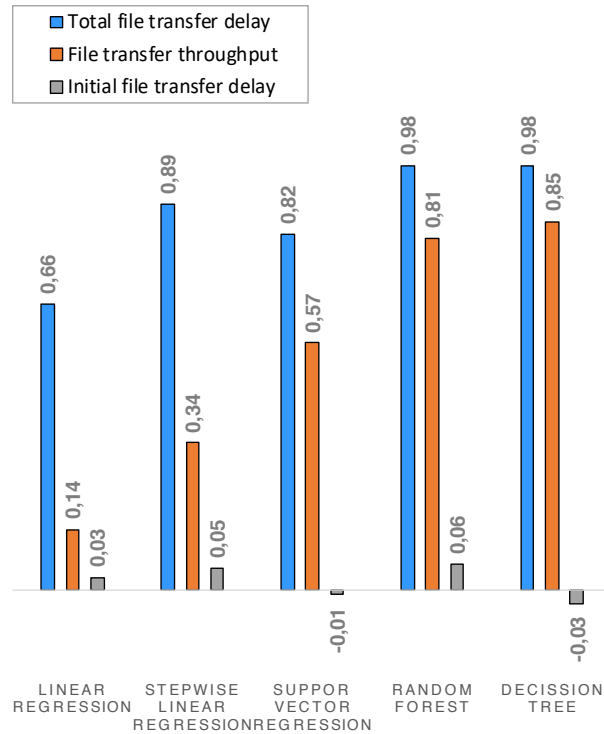

*Figure 3 Regression techniques comparison based on $R^2$*

Figure 3 shows the goodness of each of these models for the mapped KQIs, evaluated by 5-fold cross-validation. The results show that the RFR and the DTR algorithms provides the model with maximum fitness with $R^2$ values close to 1 for the *total file transfer delay (TFTD)* and the *file transfer average throughput (FTHR)*. Based on that, it can be inferred that they can be properly modelled based on the selected radio network parameters. For the *initial file transfer delay* KQI (IFTD), however, all the tested models, exhibits values $R^2$ close to zero. Unlike the previous case, this KQI cannot be modelled as a combination of radio network parameters as it could be expected: for all samples analysed the total initial transfer delay is nearly constant and completely dependent on the computational speed of the FTP server.

Table 1 shows the significance of the different low-layer metrics in the KQIs prediction performance for the DTR case. This is a measure in terms of the *predictive measure of association* coefficient. For a tree model, this coefficient estimates the level of importance of an independent variable of the model by summing the changes in the mean squared error due to splits on every variable and dividing the sum by the number of branch nodes. The lower the value is, the smallest importance it has. As it can be expected, these results show that the available bandwidth, as well as the file size, have a powerful impact on the experienced KQIs. They also show that the load is determinant in the TFTD, while the RSRP and the RSSI highly affect the FTHR. It is observed also how for the IFTD KQI all the considered inputs have quite irrelevant contributions, as it was also seen in Figure 3.

Table 1 Parameter contribution to the decision tree model

|  | BW value | Load | RSRP | RSRQ | RSSI | File size |
|---|---|---|---|---|---|---|
| TFTD | 4.029 | 0.164 | 0.024 | 0.070 | 0.040 | 8.559 |
| THR ($1*10^8$) | 0.119 | 0.033 | 0.413 | 0.038 | 0.130 | 1.165 |
| IFTD ($1*10^{-4}$) | 0.025 | 0.001 | 0.105 | 0.086 | 0.370 | 0.274 |

The approach has been further evaluated to validate how well it adapts to partial training datasets. To this effect, the samples were split into two datasets. The dataset A was composed by the data from experiments with files of 1KB, 100KB, 1MB, 10MB and 100MB. Meanwhile, the dataset B contains the measurement for experiments with file sizes of 10KB, 500KB, 5MB and 20MB sizes. Also, these datasets were split again into two groups, training (70%) and testing (30%) In Figure 4, the performance results of using dataset A to train the models while dataset B is used to test them (*partial dataset* in the figure) are compared with the results where the model is trained with the *full dataset (*A+B) and evaluated just with dataset B. In this way, in the partial dataset case the models are applied and tested for experiments (dataset B) with values of file size never included in their training (dataset A).

The metric used to estimate the performance of the models has been again the $R^2$. As expected, the results also show that the performance of the models trained with the full dataset is better that the one done with the partial one. For the full dataset case, the $R^2$ has similar values to the one shown in Figure 3. For most of the tested algorithms, the difference is not quite significant. However, for DTR and RFR, the most precise regression techniques using the full dataset, are also the most impacted by using a partial dataset (i.e. being applied for configurations different to the ones used for their training) in the TFTD KQI. As it could be expected, the performance of the decision trees and random forests algorithms highly degrades as they are mechanisms that do not properly extrapolates to values not included in their training dataset [15].

Figure 5 shows how the DTR model fits TTD and THR in a set of example measurements. For each download experiment, the measured and estimated KQI values are shown. The model used in this representation is the *full dataset* case and the evaluation is done for the group B. The Root Mean Square Error (RMSE) for the model in the complete dataset is also represented in the figure as an added shadow interval above and below of the KQI estimated values. Each lot of samples, marked in the figure with the discontinuous line, corresponds to a specific bandwidth, following, left to right, the values of 5 MHz, 10 MHz, 15 MHz and 20 MHz.

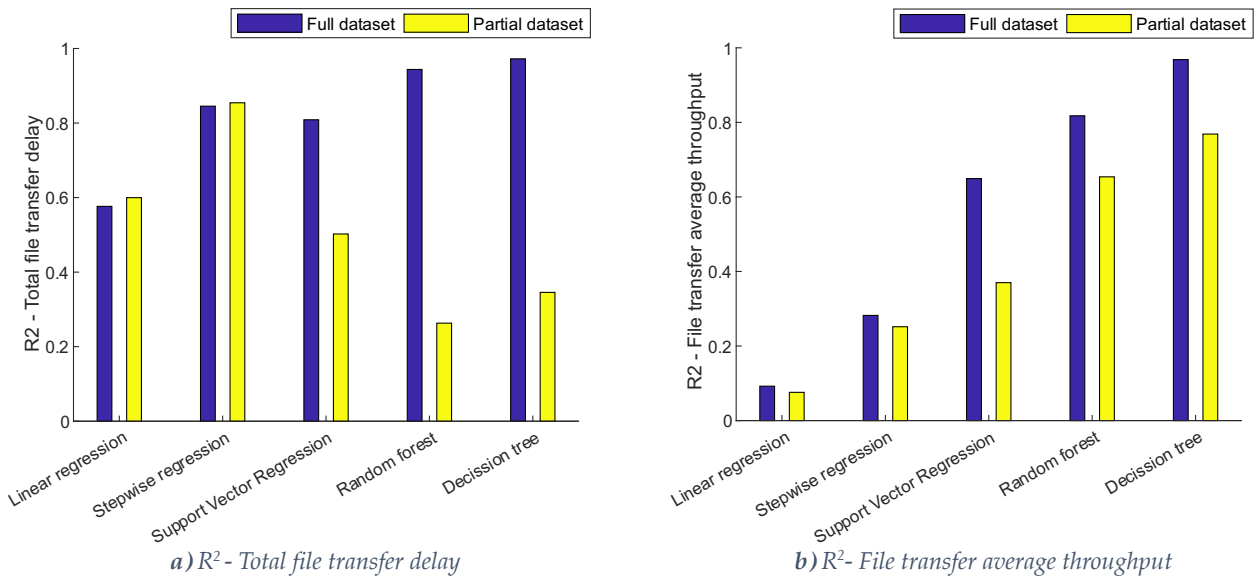

a) $R^2$ - Total file transfer delay       b) $R^2$- File transfer average throughput

Figure 4 Regression techniques comparison for full and partial data sets.

The figure allows to observe how the predicted KQIs are very close to the real ones.

## 5 CONCLUSIONS & OUTLOOK

The end-to-end approach of the upcoming 5G network management and the difficulties in collecting the service KQIs have been identified, pointing out the need of tools able to predict the service quality from metrics directly measurable by the operators. This work has proposed a novel framework and associated machine learning mechanisms allowing the translation from cellular low-layer metrics and configuration parameters to higher layer metrics (KQIs).

After defining this scheme, its implementation in a real LTE network has been described. Based on this, a comparison between different regression techniques have been performed, assessing the quality of the generated models. Their evaluation indicates good fitness levels and prediction accuracy, showing the relevance of the proposed approach.

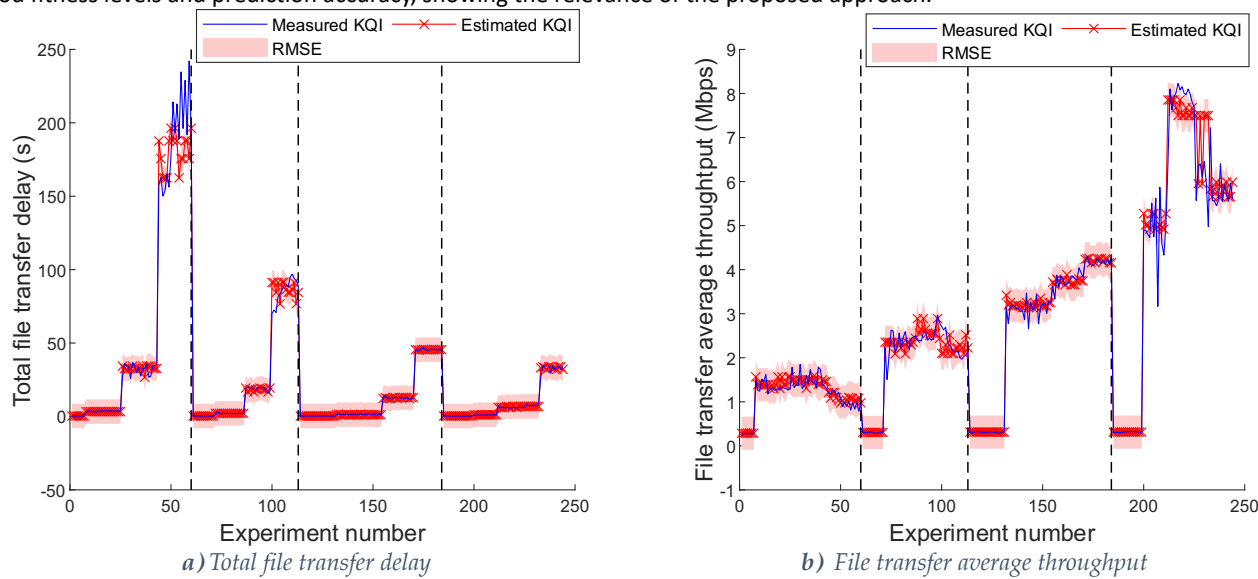

*a)* Total file transfer delay    *b)* File transfer average throughput

Figure 5 DTR model KQI prediction.

## 6 ACKNOWLEDGEMENTS

This work has been partially performed in the framework of the Horizon 2020 project ONE5G (ICT-760809) receiving funds from the European Union. The authors would like to acknowledge the contributions of their colleagues in the project, although the views expressed in this contribution are those of the authors and do not necessarily represent the project. This work has been also partially supported by the European Union's Horizon 2020 research and innovation program under grant agreement No. 644399 (MON-ROE) through the second open call project.